%% file: l3Zeepaper.tex
\documentclass[12pt,a4paper,dvips]{article}
\usepackage{a4p}
\usepackage{cite,mcite}
\usepackage{graphicx}
\usepackage{physics }
\usepackage{l3_title,ifthen}
\usepackage{rotating}
\usepackage{epsfig}
\usepackage{amsmath}
\journalname{Physics Letters B}
\date{December 2, 2002}
\preprint{2002-103}

\usepackage{Lep}
\Lep{2}

\newcommand{\grados[1]}{\ensuremath{#1^{\circ}}}

\newlength{\capindent}
\newlength{\capwidth}
\newlength{\figwidth}
\newcommand{\icaption}[2][!*!,!]{\hspace*{\capindent}%
  \begin{minipage}{\capwidth}
    \ifthenelse{\equal{#1}{!*!,!}}%
    {\caption{#2}}%
    {\caption[#1]{#2}}
  \end{minipage}}

\begin{document}
\setlength{\unitlength}{1mm}
\begin{titlepage}
\title{Study of the \boldmath{$\rm e^+e^-\rightarrow Ze^+e^-$} process
 at LEP}
\author{The L3 Collaboration}
\begin{abstract}

The cross section of the process $\rm e^+e^-\rightarrow Ze^+e^-$ is
measured with ~0.7\,fb$^{-1}$ of data collected with the L3 detector
at LEP. Decays of the Z boson into quarks and muons are considered at
centre-of-mass energies ranging from $183\GeV$ up to $209\GeV$. The measurements 
are found to agree with Standard Model predictions, achieving a precision of
about 10\% for the hadronic channel. 

\end{abstract}

\submitted

\end{titlepage}

\section*{Introduction}

The study of gauge boson production
in $\rm e^+e^-$ collisions constitutes one of the main subjects of the
scientific program carried out at LEP. Above the Z
resonance, in addition to the $s$- and $t$-channel pair-production
processes, ``single'' weak gauge bosons can also be produced via
$t$-channel processes. A common feature of this single boson production
is the emission of a virtual photon off the incoming electron or positron. 
This electron or positron remains in turn almost unscattered 
at very low polar angles and hence not detected. Particular care has to 
be paid when predicting the cross sections of these processes due to the running
of the electromagnetic coupling of the photon and the peculiarities
of the modelling of small angle scattering. The comparison of these
predictions with experimental data is made more interesting by the
fact that single boson production will constitute a copious source of
bosons at higher-energy $\rm e^+e^-$ colliders.  In addition, this 
process constitutes a significant background for the search of Standard Model 
Higgs boson or new particles predicted in physics beyond the Standard Model. 
The ``single W'' production is extensively studied at LEP\cite{SingleZ,SingleW} 
and this Letter concentrates on ``single Z'' production. Results at lower 
centre-of-mass energies were previously reported\cite{SingleZ,opal}. 

\begin{figure}[p]
 \begin{center}
     \mbox{\includegraphics[width=\figwidth]{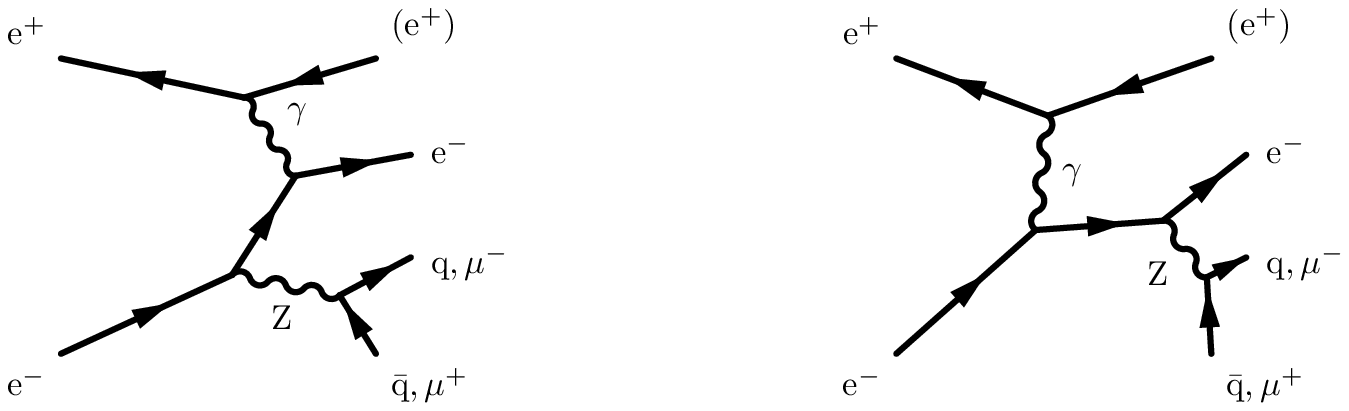}} 
   \icaption{Main diagrams contributing to the ``single Z'' production.
     \label{fig:1}}
 \end{center}
\end{figure}

Figure~\ref{fig:1} presents two Feynman diagrams for the single Z
production, followed by the decay of the Z into a quark-antiquark
or a muon-antimuon pair. A distinctive feature of this process 
is the photon-electron scattering, reminiscent of the Compton scattering. 
These diagrams are only an example of the 48 diagrams contributing to the 
$\rm e^+e^- \rightarrow q\bar{q}e^+e^-$ and $\rm e^+e^- \rightarrow 
\mu^+\mu^-e^+e^-$ final state processes.  The single Z signal is defined 
starting from this full set of diagrams. QCD contributions 
from two-photon physics with $\rm e^+e^- \rightarrow q\bar{q}e^+e^-$ 
final state are not considered. The definition requires the final state 
fermions to satisfy the kinematical cuts:
\begin{equation}
m_{\rm f\bar{f}} > 60\GeV,\,\,\, \theta_{unscattered}<\grados[12],\,\,\, 
\grados[60] < \theta _{scattered} < \grados[168],\,\,\, E_{scattered} > 3.0 \GeV,
\end{equation}
where $m_{\rm f\bar{f}}$ refers to the invariant mass of the
produced quark-antiquark or muon-antimuon pair, $\theta_{unscattered}$
is the polar angle at which the electron\footnote{The word
``electron'' is used for both electrons and positrons.}  closest
to the beam line is emitted, 
$\theta _{scattered}$ and $E_{scattered}$ are respectively the polar angle  
with respect to its incoming direction and the energy of the electron scattered 
at the largest polar angle.

These criteria largely enhance the contribution of diagrams similar to
those in Figure~\ref{fig:1} over the remaining phase space of the $\rm
e^+e^- \rightarrow q\bar{q}e^+e^-$ and  $\rm
e^+e^- \rightarrow \mu^+\mu^-e^+e^-$ processes and correspond to predicted
 cross sections at a centre-of-mass energy $\sqrt{s}=200\GeV$  of about
0.6\,pb for the hadron channel and of about 0.04\,pb for the muon one. The most severe 
backgrounds for the detection of the single Z production at LEP are the
$\rm e^+e^- \rightarrow q\bar{q}(\gamma)$ and the $\rm e^+e^-\rightarrow \mu^+\mu^-(\gamma)$ 
processes, for the hadron and muon channels respectively.

This Letter describes the selection of  $\rm e^+e^- \rightarrow Ze^+e^-
\rightarrow q\bar{q}e^+e^-$ and   $\rm e^+e^- \rightarrow Ze^+e^-
\rightarrow \mu^+\mu^-e^+e^-$  events in the data sample collected by the
L3 detector\cite{l3} at LEP and the measurement of the cross section of these processes.

\section*{Data and Monte Carlo samples}

This analysis is based on $675.5\,\rm pb^{-1}$ of integrated
luminosity collected at \mbox{$\sqrt{s}=182.7-209.0\GeV$}. For the investigation of 
the $\rm e^+e^- \rightarrow Ze^+e^- \rightarrow q\bar{q}e^+e^-$ channel, this sample
is divided into eight different energy bins whose corresponding average
$\sqrt{s}$ values and integrated luminosities are reported in 
Table~\ref{tab:1}.

\begin{table} [htn]
  \begin{center}
    \begin{tabular}{|c|cccccccc|}
      \hline
      \rule {0pt}{12pt} $\rm{\sqrt{s}\,[{GeV}]}$ & 182.7 & 188.6 & 191.6 & 195.5 & 199.5 & 201.7 & 204.9 & 206.6 \\
      \hline
      \phantom{0}$\cal L$\,[pb$^{-1}$] & \phantom{0}55.1 & 176.0 & \phantom{0}29.4 & \phantom{0}83.0 & \phantom{0}80.8 & \phantom{0}36.7 & \phantom{0}76.6 & 137.9 \\
      \hline
      \end{tabular}
      \icaption[tab:1]{The average centre-of-mass energies and the corresponding
      integrated luminosities of the data sample used in this study.\label{tab:1}}
    \end{center}
\end{table}  

The signal process is modelled with the WPHACT Monte Carlo
program~\cite{wphact}. The GRC4F~\cite{grace} event generator is used
for systematic checks. Events are generated in a phase space broader
than the one defined by the criteria (1). Those events who do not satisfy
these criteria are considered as background. The $\rm
e^+e^-\rightarrow q\bar{q}(\gamma)$, $\rm e^+e^-\rightarrow
\mu^+\mu^-(\gamma)$ and $\rm e^+e^-\rightarrow\tau^-\tau^+(\gamma)$ processes
are simulated with the KK2f~\cite{kk2f} Monte Carlo generator, the
$\rm e^+e^-\rightarrow ZZ$ process with PYTHIA~\cite{pythia}, and the $\rm
e^+e^-\rightarrow W^+W^-$ process, with the exception of the $\rm
q\bar{q}'e\nu$ final state, with KORALW~\cite{koralw}.
EXCALIBUR~\cite{excalibur} is used to simulate the $\rm q\bar{q}'e\nu$
and other four-fermion final states. Hadron and lepton production
in two-photon interactions are modelled with PHOJET~\cite{phojet} and
DIAG36~\cite{diag36}, respectively. The generated events are passed
through the L3 detector simulation program~\cite{l3sim}. Time
dependent detector inefficiencies, as monitored during the data taking
period, are also simulated.
 
\section*{Event selection}

\subsection*{\boldmath{$\rm e^+e^- \rightarrow Ze^+e^-
\rightarrow q\bar{q}e^+e^-$} channel}

The selection of events in the $\rm e^+e^- \rightarrow Ze^+e^-
\rightarrow q\bar{q}e^+e^-$ channel proceeds from high multiplicity
events with at least one electron identified in the BGO
electromagnetic calorimeter and in the central tracker with an energy
above 3~GeV. Electron isolation criteria are applied. These are based
on the energy deposition and track multiplicity around the electron
candidate.

To strongly suppress the contribution from the high cross section
background processes, the signal topology is enforced requiring events
with a reconstructed invariant mass of the hadronic system, stemming
from a Z boson, between 50 and 130\,\GeV, a visible energy of at least
$0.40\sqrt{s}$ and a missing momentum, due to the undetected electron,
of at least $0.24\sqrt{s}$.  These quantities are computed from
charged tracks, calorimetric clusters and possible muons. After these
selection criteria, 1551 events are selected in the full data
sample. From Monte Carlo, $1551 \pm 4$ events are expected, out of
which $208 \pm 1$ are signal events, selected with an efficiency of
54\%. Most of the background arises from the $\rm e^+e^-\rightarrow
q\bar{q}'e\nu$ (58\%), $\rm e^+e^-\rightarrow q\bar{q}(\gamma)$ (19\%)
and $\rm e^+e^-\rightarrow W^+W^-$ (11\%) processes.

The particular signature  of an electron undetected at
low angle and the other scattered in the detector, allows to reject a
large fraction of the background by considering two powerful kinematic
variables: the product of the charge, $q$, of the detected electron
and the cosine of its polar angle measured with respect to the
direction of the incoming electron, $\cos{\theta}$, and the product of
$q$ and the polar angle of the direction of the missing momentum,
$\cos{\theta\hspace{-0.5em}/}$. Two selection criteria are applied:
\begin{displaymath}
q\times \cos{\theta} > -0.5\,\,\,{\rm and}\,\,\, q\times\cos{\theta\hspace{-0.5em}/}>0.94.
\end{displaymath}
Distributions of these variables are presented in Figure~\ref{fig:2}.
In addition, events are forced into two jets by means of the DURHAM
algorithm~\cite{durham}, and the opening angle between the two jets in the plane transverse to the beam direction  is required to
exceed $\grados[150]$. The selected electrons are not considered when forming those jets. Background events are further rejected by
tightening the electron isolation criteria.
Table~\ref{tab:2} summarises the yield of this event selection.

\begin{table} [htn]
  \begin{center}
    \begin{tabular}{|c|c|c|c|c||c|c|c|}
      \hline
      \rule {0pt}{12pt} $\sqrt{s}\rm\,[GeV]$ & $\varepsilon$\,[\%] &
      $N_{\rm Data}$ & $N_{\rm MC}$ & $N_{\rm Sign}$ &$N_{\rm{q\qbar(\gamma)}}$ & 
      $N_{\rm{q\bar{q}'e\nu}}$ &
      $N_{\rm two-phot}$\\
       \hline
       182.7 & 42.3 &            16 & $16.0 \pm 0.5 $            & $ 12.0 \pm 0.2$            & 3.2 & 0.4  & 0.2\\   
       188.6 & 42.7 & 53            & $52.4 \pm 1.2 $            & $ 40.3 \pm 0.6$            & 9.5 & 0.5  & 1.0\\ 
       191.6 & 43.0 & \phantom{0}9  & $\phantom{0}8.7  \pm 0.3 $ & $ \phantom{0}6.9 \pm  0.2$ & 1.3 & 0.0  & 0.3\\
       195.5 & 45.0 & 19            & $26.5 \pm 0.6 $            & $ 21.2 \pm 0.3$            & 3.6 & 0.0  & 1.1\\
       199.5 & 45.2 & 18            & $27.1 \pm 0.7 $            & $ 21.3 \pm 0.3$            & 3.8 & 0.4  & 1.2\\
       201.7 & 44.0 & 16            & $12.2 \pm 0.5 $            & $ \phantom{0}9.5  \pm 0.2$ & 1.6 & 0.3  & 0.5\\
       204.9 & 43.3 & 24            & $24.6 \pm 0.4 $            & $ 19.9 \pm 0.3$            & 2.9 & 0.2  & 1.1\\
       206.6 & 44.6 & 47            & $46.2 \pm 0.7 $            & $ 37.2 \pm 0.6$            & 5.4 & 0.3  & 2.2\\
       \hline
      \end{tabular}
      \icaption[tab:2]{Yield of the $\rm e^+e^- \rightarrow Ze^+e^-
        \rightarrow q\bar{q}e^+e^-$ event selection at the different
        centre-of-mass energies. The signal efficiency, $\varepsilon$,
        is listed together with the number of observed, $N_{\rm Data}$,
        and total expected, $N_{\rm MC}$, events. The
        expected number of signal events, $N_{\rm Sign}$, is given
        together with details of the most important residual
        backgrounds, respectively indicated with
        $N_{\rm{q\qbar(\gamma)}}$, $N_{\rm{ q\bar{q}'e\nu}}$ and $N_{\rm
          two-phot}$ for the processes $\rm e^+e^-\rightarrow
        q\qbar(\gamma)$, $\rm e^+e^-\rightarrow q\bar{q}'e\nu$ and
        hadron production in two-photon interactions.\label{tab:2}}
    \end{center}
\end{table}     

\subsection*{\boldmath{$\rm e^+e^- \rightarrow Ze^+e^-
\rightarrow \mu^+\mu^-e^+e^-$} channel}

Candidates for the $\rm e^+e^- \rightarrow Ze^+e^-
\rightarrow \mu^+\mu^-e^+e^-$ process are selected by first requiring
low multiplicity events with three tracks in the central tracker,
corresponding to one electron with energy above $3\GeV$ and two muons, reconstructed in the muon
spectrometer with momenta above 18\,\GeV. A kinematic fit is then
applied which requires momentum conservation in the plane transverse to
the beam axis. The reconstructed invariant mass of the two muons should lie between
55 and 145\,\GeV. Finally, three additional selection criteria are
applied:
\begin{displaymath}
-0.50 < q\times \cos{\theta}<0.93,\,\,\, q\times\cos{\theta\hspace{-0.5em}/}>0.50\,\,\,{\rm and}\,\,\,q\times \cos{\rm\theta_Z}< 0.40,
\end{displaymath}
where $\cos{\rm\theta_Z}$ is the polar angle of the Z boson as
reconstructed from the two muons. These criteria select 9 data events and 
$6.6\pm 0.1$ expected events from signal Monte Carlo with an
efficiency of 22\%. Background expectations amount to $1.5\pm 0.1$
events, coming in equal parts from muon-pair production in two-photon interactions, the
$\rm e^+e^-\rightarrow \mu^+\mu^-(\gamma)$ process, and 
$\rm e^+e^- \rightarrow \mu^+\mu^-e^+e^-$ events generated with WPHACT that 
do not pass the signal definition criteria.

\section*{Results}

\begin{figure}[p]
 \begin{center}
   \begin{tabular}{cc}
   \mbox{\includegraphics[width=.5\figwidth]{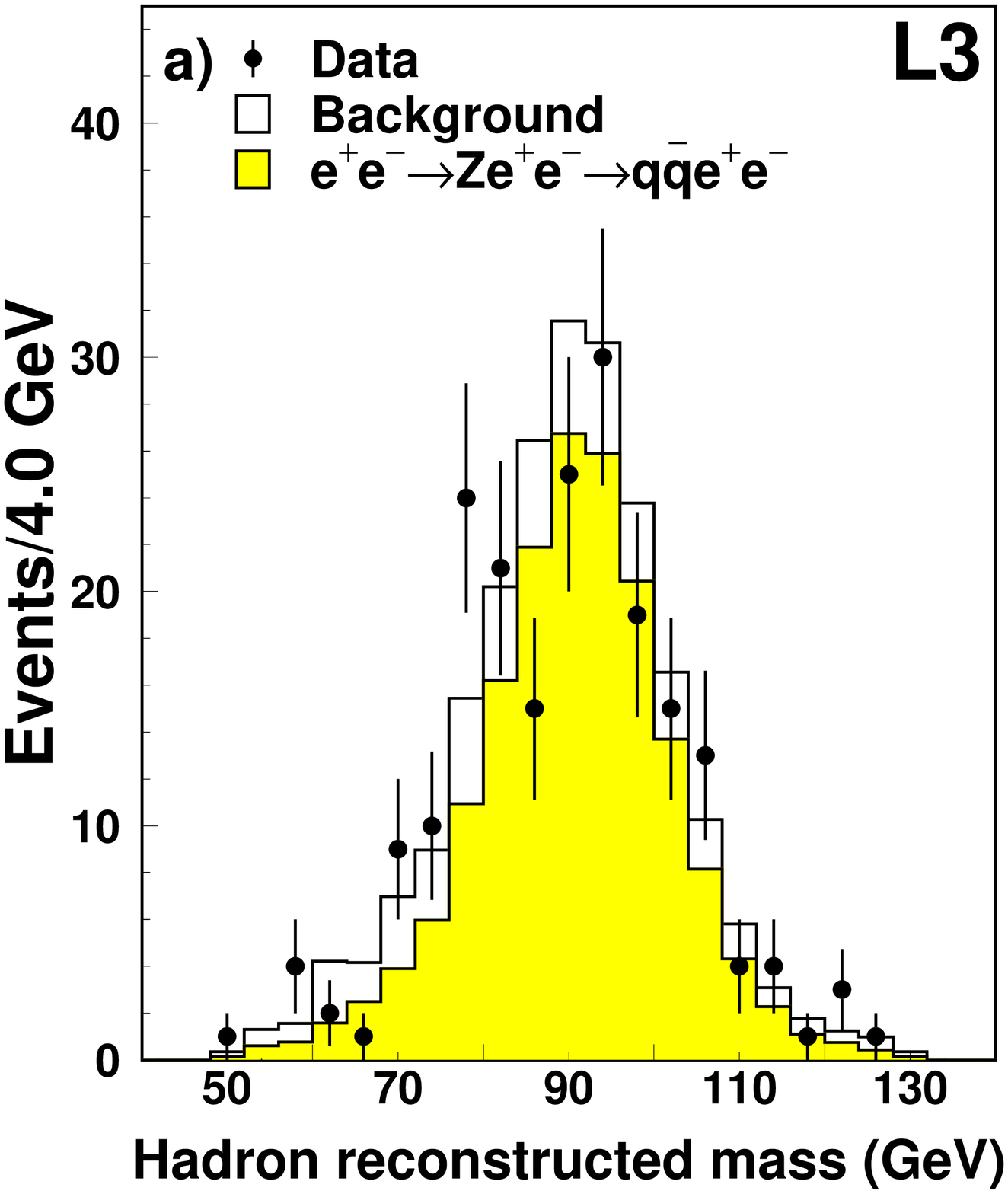}} &
   \mbox{\includegraphics[width=.5\figwidth]{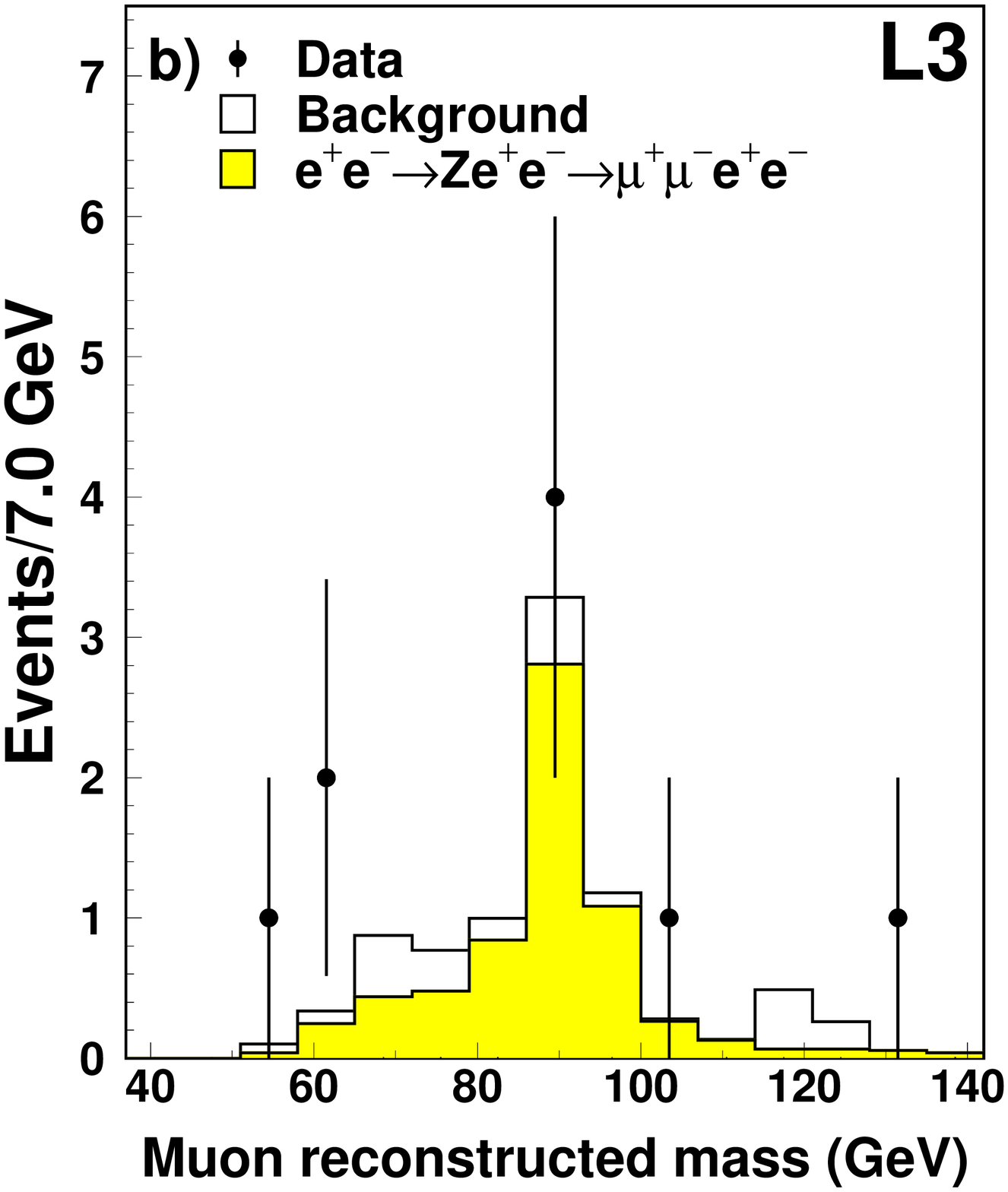}} \\
 \end{tabular}
   \icaption{Distribution of the reconstructed invariant mass of a) the hadron 
system and b) the muon system for data, signal, and background Monte Carlo events.
     \label{fig:3}}
 \end{center}
\end{figure}

Figure~\ref{fig:3}a presents the distribution of the invariant mass of
the hadronic system after applying all selection criteria of the $\rm e^+e^-
\rightarrow Ze^+e^- \rightarrow q\bar{q}e^+e^-$ channel. A
large signal peaking around the mass of the Z boson is observed. The single Z
cross section at each value of $\sqrt{s}$ is determined from a
maximum-likelihood fit to the distribution of this variable. The
results are listed in Table~\ref{tab:3}, together with the predictions
of the WPHACT Monte Carlo. A good agreement is observed.

\begin{table} [htn]
  \begin{center}
    \begin{tabular}{|c|c|c|}
      \hline
      \rule {0pt}{12pt} $\sqrt{s}$\,[GeV]    & $\sigma^{\rm
      Measured}$\,[pb] & $\sigma^{\rm Expected}$\,[pb] \\
      \hline
      \rule {0pt}{12pt}182.7 & $0.51^{+0.19}_{-0.16}\pm 0.03$ & 0.51\\
      \rule {0pt}{12pt}188.6 & $0.54^{+0.10}_{-0.09}\pm 0.03$ & 0.54\\
      \rule {0pt}{12pt}191.6 & $0.60^{+0.26}_{-0.21}\pm 0.04$ & 0.55\\
      \rule {0pt}{12pt}195.5 & $0.40^{+0.13}_{-0.11}\pm 0.02$ & 0.56\\
      \rule {0pt}{12pt}199.5 & $0.33^{+0.12}_{-0.10}\pm 0.02$ & 0.58\\
      \rule {0pt}{12pt}201.7 & $0.81^{+0.26}_{-0.22}\pm 0.05$ & 0.59\\
      \rule {0pt}{12pt}204.9 & $0.55^{+0.16}_{-0.14}\pm 0.03$ & 0.60\\ 
      \rule {0pt}{12pt}206.6 & $0.59^{+0.12}_{-0.10}\pm 0.03$ & 0.61\\
      \hline
      \end{tabular}
      \icaption[tab:3]{Measured and expected cross sections for the
        $\rm e^+e^- \rightarrow Ze^+e^- \rightarrow q\bar{q}e^+e^-$ process at
        the different centre-of-mass energies. The first uncertainties are
        statistical and the second systematic. Expectations are
        calculated with the WPHACT Monte Carlo program.\label{tab:3}}
    \end{center}
\end{table}     

The invariant mass of muon pairs from the  $\rm e^+e^- \rightarrow Ze^+e^-
\rightarrow \mu^+\mu^-e^+e^-$ selected events is shown in
Figure~\ref{fig:3}b. The cross section of this process is determined with a fit to the invariant mass distribution, over the full data sample, as:
\begin{displaymath}
\sigma(\rm e^+e^- \rightarrow Ze^+e^-
\rightarrow \mu^+\mu^-e^+e^-) = 0.043 ^{+0.013} _{-0.013} \pm 0.003\,pb
\,\,\,\,\, \left( \sigma^{SM} = 0.044\,pb \right),
\end{displaymath}

where the first uncertainty is statistical and the second systematic.
This measurement agrees with the Standard Model prediction $\rm{\sigma^{SM}}$ 
reported in parenthesis, which is calculated with the WPHACT program as the 
luminosity weighted average cross section over the different centre-of-mass energies.

\begin{figure}[p]
 \begin{center}
   \begin{tabular}{cc}
   \mbox{\includegraphics[width=.5\figwidth]{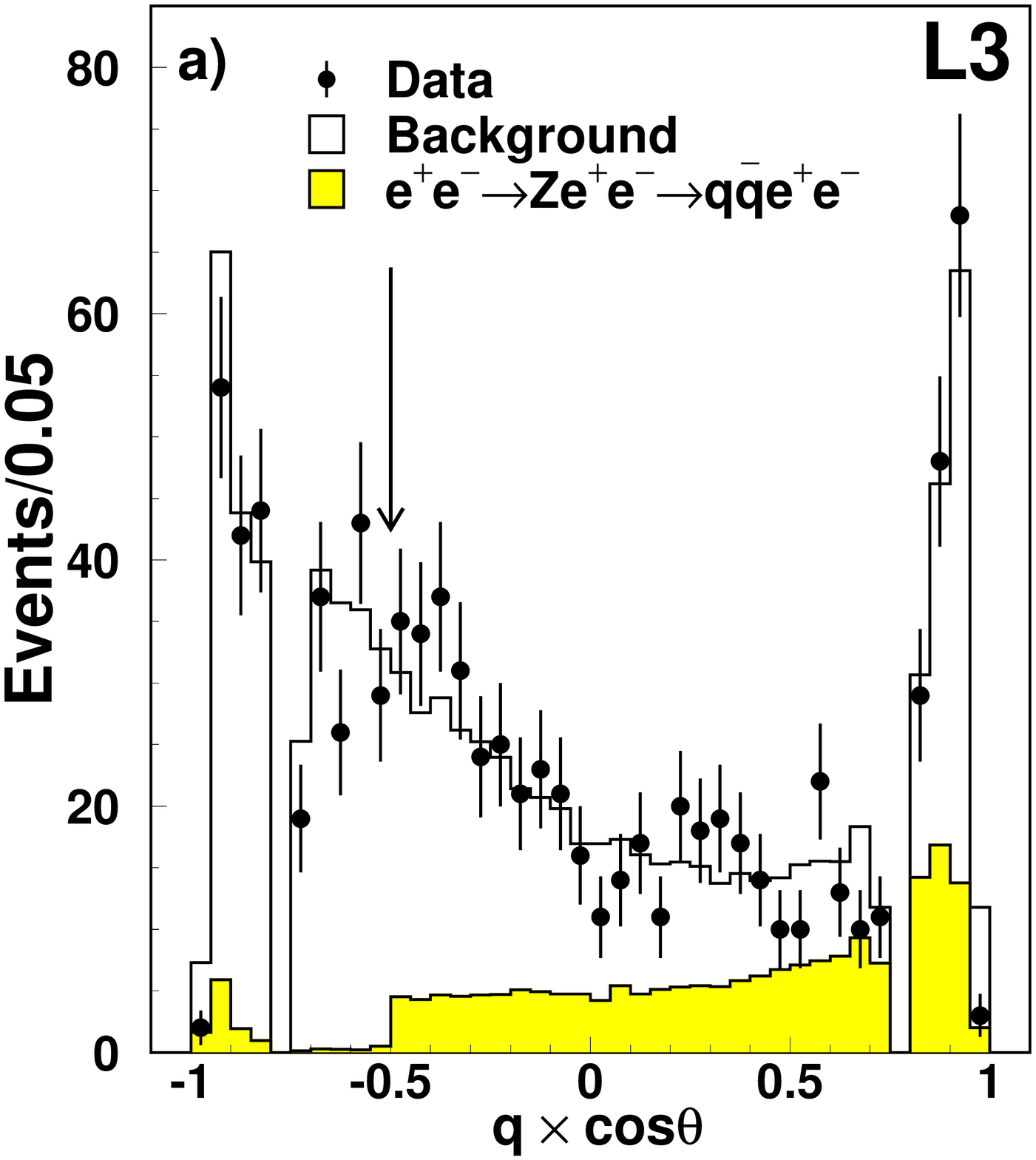}} &
   \mbox{\includegraphics[width=.5\figwidth]{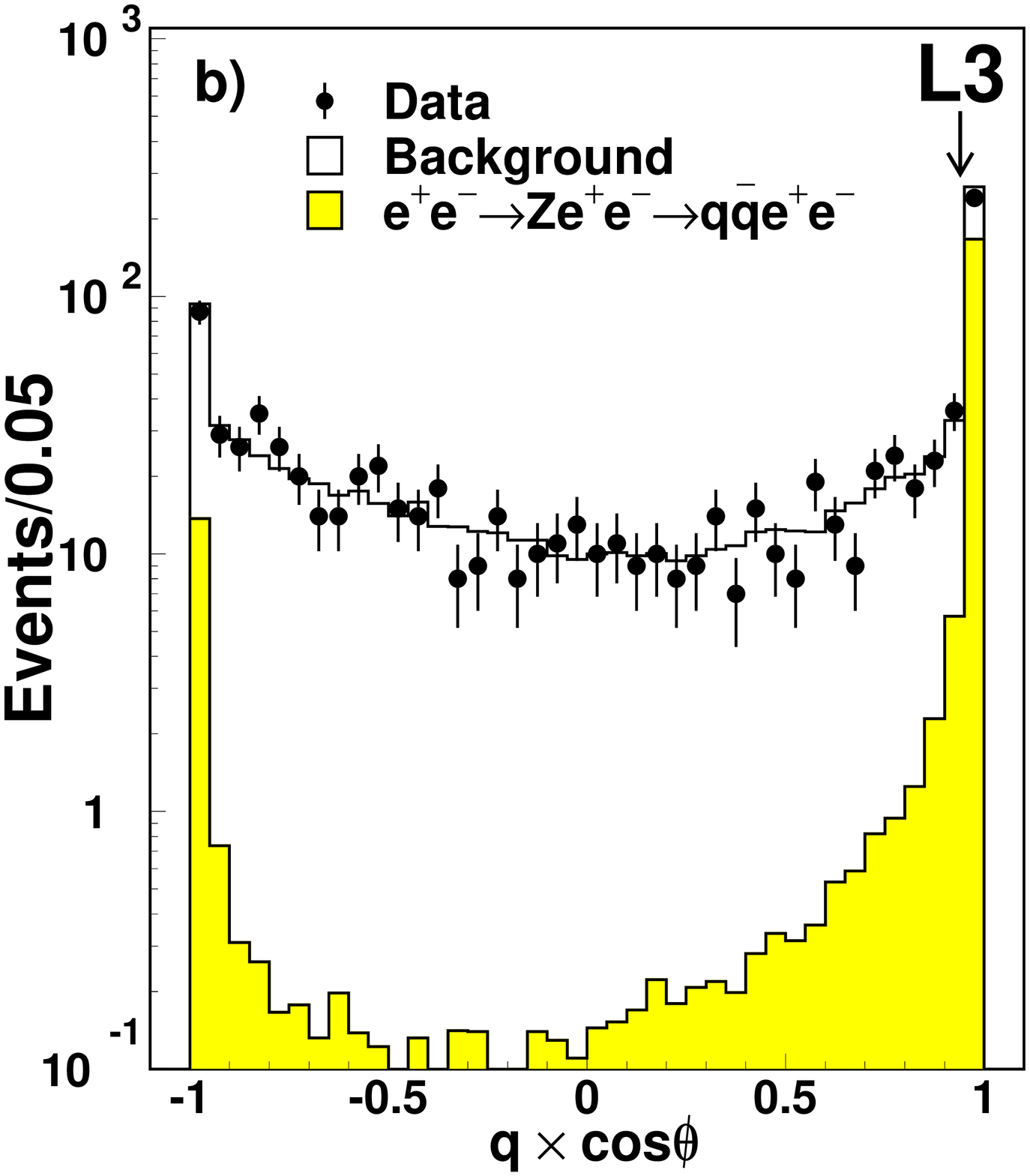}} \\
 \end{tabular}
 \icaption{Distributions for data, signal and background Monte Carlo
 of the product of the charge of the detected electron and a) the 
 cosine of its polar angle and b) the cosine of the polar angle of the
 missing momentum. The arrows show the position of the applied
 cuts. All other selection criteria but those on these two variables are applied. 
 Signal events around $-1$ correspond to charge confusion in the central tracker.
The sharp edge of the signal distribution in a) at $-0.5$ follows from the
signal definition criterion $\theta_{scattered} > 60^\circ$; moreover, the
depletion around $\pm0.7$ in data and Monte Carlo is due to the absence of    
the BGO calorimeter in this angular region.   
\label{fig:2}}
 \end{center}
\end{figure}

\begin{figure}[p]
 \begin{center}
     \mbox{\includegraphics[width=\figwidth]{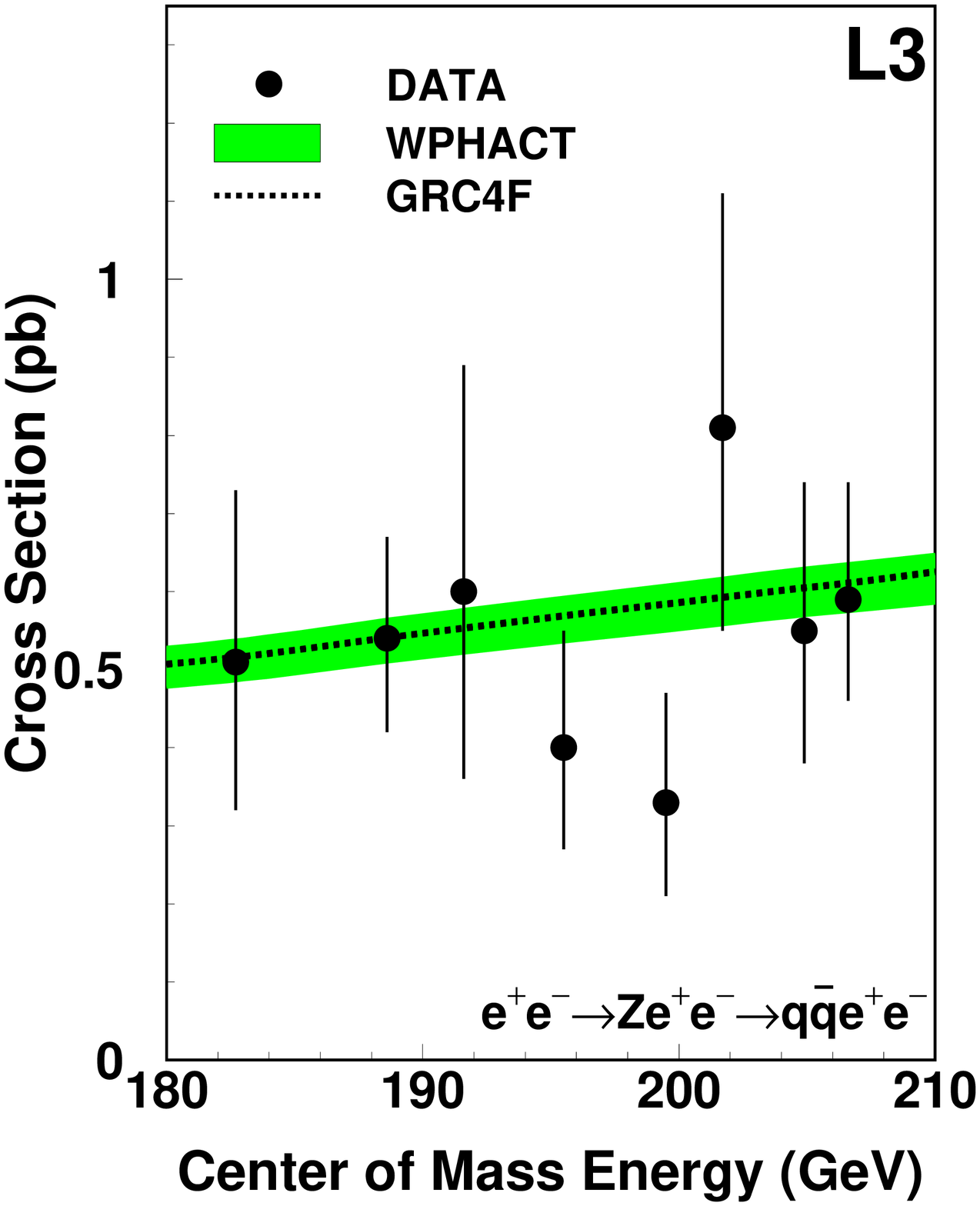}} 
   \icaption{Measurements of the cross section of the $\rm e^+e^-
   \rightarrow Z  e^+e^- \rightarrow q\bar{q}e^+e^-$ process as a function of the
   centre-of-mass energy. The WPHACT predictions are assigned an
   uncertainty of 5\%. As reference, a line indicates the GRC4F expectations.
     \label{fig:4}}
 \end{center}
\end{figure}

Several possible sources of systematic uncertainty are considered and
their effects on the measured cross sections are listed in
Table~\ref{tab:4}. First, detector effects and the accuracy of the
Monte Carlo simulations are investigated by varying the energy scale of
the calorimeters, the amount of charge confusion in the tracker,
visible for instance in Figure~\ref{fig:2} as the signal enhancement on the left
side, and the selection criteria. The impact of the signal modelling on the final 
efficiencies is studied by using the GRC4F Monte Carlo program instead of the
WPHACT event generator to derive the signal efficiencies. The expected
cross sections of the background processes for the $\rm e^+e^-
\rightarrow Ze^+e^- \rightarrow q\bar{q}e^+e^-$ channel are varied by
5\% for $\rm e^+e^- \rightarrow q\bar{q}(\gamma)$, 10\% for $\rm e^+e^- 
\rightarrow q\bar{q}'e\nu$, 1\% for $\rm e^+e^- \rightarrow W^+W^-$,
and 50\% for hadron production in two-photon interactions. The cross sections 
of the background
processes for the $\rm e^+e^- \rightarrow Ze^+e^- \rightarrow
\mu^+\mu^-e^+e^-$ channel are varied by 2\% for the $\rm e^+e^-
\rightarrow \mu^+\mu^-(\gamma)$ channel, 10\% for the WPHACT $\rm
e^+e^- \rightarrow \mu^+\mu^-e^+e^-$ events that do not pass the
signal definition and 25\% for muon-pair production in two-photon
interactions. Finally, the effects of the limited background and signal
Monte Carlo statistics are considered.
\begin{table}[h]
  \begin{center}
    \begin{tabular}{|l|r @{.} l|r @{.} l|}
      \hline
      Source & \multicolumn{4}{|c|}{Systematic uncertainty} \\
      \cline{2-5}
       & \multicolumn{2}{|c|}{
         $\rm e^+e^- \rightarrow Z  e^+e^-\rightarrow q\bar{q} e^+e^-$} 
       & \multicolumn{2}{|c|}{ $\rm e^+e^- \rightarrow Ze^+e^- \rightarrow \mu^+\mu^-e^+e^-$} \\
      \hline
      Energy scale                       & \phantom{000000000l}2&3\% &\phantom{00000000000}6&3\% \\
      Charge confusion                   & 0&8\% & $<$0&1\% \\
      Selection procedure                & 4&0\% & 1&9\% \\
      Signal modelling                   & 1&2\% & $<$0&1\% \\
      Background modelling               & 1&0\% & 2&9\% \\
      Background Monte Carlo statistics  & 2&8\% & 1&8\% \\
      Signal Monte Carlo statistics      & 1&6\% & 2&2\% \\
      \hline
      Total                              & 5&9\% & 7&7\% \\
      \hline
      \end{tabular}
      \icaption[tab:4]{Sources of systematic uncertainties.\label{tab:4}}
    \end{center}
\end{table}      
Figure~\ref{fig:4} compares the results of the measurement of the cross section
of the process  $\rm e^+e^- \rightarrow Z  e^+e^-\rightarrow q\bar{q} e^+e^- $
with both the WPHACT and the GRC4F predictions. A  good
agreement is observed. This agreement is quantified by extracting the
ratio $R$ between the measured cross sections $\sigma^{\rm Measured}$
and the  WPHACT predictions $\sigma^{\rm Expected}$:
\begin{displaymath}
  R = {\sigma^{\rm Measured} \over \sigma^{\rm Expected}}= 0.88 \pm 0.08 \pm 0.06,
\end{displaymath}
where the first uncertainty is statistical and the second systematic.

In conclusion, the process $\rm e^+e^- \rightarrow Z e^+e^-$  has been 
observed at LEP for decays of the Z boson into both hadrons and muons. 
The measured cross sections have been compared with the Standard
Model predictions, and were found in agreement with an experimental accuracy 
of about 10\% for decays of the Z boson into hadrons.\

\newpage

%
%

\input namelist261.tex


%
%

\end{document}

%% file: namelist261.tex
\typeout{   }     
\typeout{Using author list for paper 261 -  }
\typeout{$Modified: Jul 15 2001 by smele $}
\typeout{!!!!  This should only be used with document option a4p!!!!}
\typeout{   }
%
%
%
%
%
%

\newcount\tutecount  \tutecount=0
\def\tutenum#1{\global\advance\tutecount by 1 \xdef#1{\the\tutecount}}
\def\tute#1{$^{#1}$}
\tutenum\aachen            
\tutenum\nikhef            
\tutenum\mich              
\tutenum\lapp              
\tutenum\basel             
\tutenum\lsu               
\tutenum\beijing           
\tutenum\bologna           
\tutenum\tata              
\tutenum\ne                
\tutenum\bucharest         
\tutenum\budapest          
\tutenum\mit               
\tutenum\panjab            
\tutenum\debrecen          
\tutenum\dublin            
\tutenum\florence          
\tutenum\cern              
\tutenum\wl                
\tutenum\geneva            
\tutenum\hefei             
\tutenum\lausanne          
\tutenum\lyon              
\tutenum\madrid            
\tutenum\florida           
\tutenum\milan             
\tutenum\moscow            
\tutenum\naples            
\tutenum\cyprus            
\tutenum\nymegen           
\tutenum\caltech           
\tutenum\perugia           
\tutenum\peters            
\tutenum\cmu               
\tutenum\potenza           
\tutenum\prince            
\tutenum\riverside         
\tutenum\rome              
\tutenum\salerno           
\tutenum\ucsd              
\tutenum\sofia             
\tutenum\korea             
\tutenum\purdue            
\tutenum\psinst            
\tutenum\zeuthen           
\tutenum\eth               
\tutenum\hamburg           
\tutenum\taiwan            
\tutenum\tsinghua          

{
\parskip=0pt
\noindent
{\bf The L3 Collaboration:}
\ifx\selectfont\undefined
 \baselineskip=10.8pt
 \baselineskip\baselinestretch\baselineskip
 \normalbaselineskip\baselineskip
 \ixpt
\else
 \fontsize{9}{10.8pt}\selectfont
\fi
\medskip
\tolerance=10000
\hbadness=5000
\raggedright
\hsize=162truemm\hoffset=0mm
\def\r{\rlap,}
\noindent

P.Achard\r\tute\geneva\ 
O.Adriani\r\tute{\florence}\ 
M.Aguilar-Benitez\r\tute\madrid\ 
J.Alcaraz\r\tute{\madrid,\cern}\ 
G.Alemanni\r\tute\lausanne\
J.Allaby\r\tute\cern\
A.Aloisio\r\tute\naples\ 
M.G.Alviggi\r\tute\naples\
H.Anderhub\r\tute\eth\ 
V.P.Andreev\r\tute{\lsu,\peters}\
% ...
\newpage
\begin{list}{A}{\itemsep=0pt plus 0pt minus 0pt\parsep=0pt plus 0pt minus 0pt
                \topsep=0pt plus 0pt minus 0pt}
\item[\aachen]
 III. Physikalisches Institut, RWTH, D-52056 Aachen, Germany$^{\S}$
\item[\nikhef] National Institute for High Energy Physics, NIKHEF, 
     and University of Amsterdam, NL-1009 DB Amsterdam, The Netherlands
\item[\mich] University of Michigan, Ann Arbor, MI 48109, USA
\item[\lapp] Laboratoire d'Annecy-le-Vieux de Physique des Particules, 
     LAPP,IN2P3-CNRS, BP 110, F-74941 Annecy-le-Vieux CEDEX, France
\item[\basel] Institute of Physics, University of Basel, CH-4056 Basel,
     Switzerland
\item[\lsu] Louisiana State University, Baton Rouge, LA 70803, USA
\item[\beijing] Institute of High Energy Physics, IHEP, 
  100039 Beijing, China$^{\triangle}$ 
\item[\bologna] University of Bologna and INFN-Sezione di Bologna, 
     I-40126 Bologna, Italy
\item[\tata] Tata Institute of Fundamental Research, Mumbai (Bombay) 400 005, India
\item[\ne] Northeastern University, Boston, MA 02115, USA
\item[\bucharest] Institute of Atomic Physics and University of Bucharest,
     R-76900 Bucharest, Romania
\item[\budapest] Central Research Institute for Physics of the 
     Hungarian Academy of Sciences, H-1525 Budapest 114, Hungary$^{\ddag}$
\item[\mit] Massachusetts Institute of Technology, Cambridge, MA 02139, USA
\item[\panjab] Panjab University, Chandigarh 160 014, India.
\item[\debrecen] KLTE-ATOMKI, H-4010 Debrecen, Hungary$^\P$
\item[\dublin] Department of Experimental Physics,
  University College Dublin, Belfield, Dublin 4, Ireland
\item[\florence] INFN Sezione di Firenze and University of Florence, 
     I-50125 Florence, Italy
\item[\cern] European Laboratory for Particle Physics, CERN, 
     CH-1211 Geneva 23, Switzerland
\item[\wl] World Laboratory, FBLJA  Project, CH-1211 Geneva 23, Switzerland
\item[\geneva] University of Geneva, CH-1211 Geneva 4, Switzerland
\item[\hefei] Chinese University of Science and Technology, USTC,
      Hefei, Anhui 230 029, China$^{\triangle}$
\item[\lausanne] University of Lausanne, CH-1015 Lausanne, Switzerland
\item[\lyon] Institut de Physique Nucl\'eaire de Lyon, 
     IN2P3-CNRS,Universit\'e Claude Bernard, 
     F-69622 Villeurbanne, France
\item[\madrid] Centro de Investigaciones Energ{\'e}ticas, 
     Medioambientales y Tecnol\'ogicas, CIEMAT, E-28040 Madrid,
     Spain${\flat}$ 
\item[\florida] Florida Institute of Technology, Melbourne, FL 32901, USA
\item[\milan] INFN-Sezione di Milano, I-20133 Milan, Italy
\item[\moscow] Institute of Theoretical and Experimental Physics, ITEP, 
     Moscow, Russia
\item[\naples] INFN-Sezione di Napoli and University of Naples, 
     I-80125 Naples, Italy
\item[\cyprus] Department of Physics, University of Cyprus,
     Nicosia, Cyprus
\item[\nymegen] University of Nijmegen and NIKHEF, 
     NL-6525 ED Nijmegen, The Netherlands
\item[\caltech] California Institute of Technology, Pasadena, CA 91125, USA
\item[\perugia] INFN-Sezione di Perugia and Universit\`a Degli 
     Studi di Perugia, I-06100 Perugia, Italy   
\item[\peters] Nuclear Physics Institute, St. Petersburg, Russia
\item[\cmu] Carnegie Mellon University, Pittsburgh, PA 15213, USA
\item[\potenza] INFN-Sezione di Napoli and University of Potenza, 
     I-85100 Potenza, Italy
\item[\prince] Princeton University, Princeton, NJ 08544, USA
\item[\riverside] University of Californa, Riverside, CA 92521, USA
\item[\rome] INFN-Sezione di Roma and University of Rome, ``La Sapienza",
     I-00185 Rome, Italy
\item[\salerno] University and INFN, Salerno, I-84100 Salerno, Italy
\item[\ucsd] University of California, San Diego, CA 92093, USA
\item[\sofia] Bulgarian Academy of Sciences, Central Lab.~of 
     Mechatronics and Instrumentation, BU-1113 Sofia, Bulgaria
\item[\korea]  The Center for High Energy Physics, 
     Kyungpook National University, 702-701 Taegu, Republic of Korea
\item[\purdue] Purdue University, West Lafayette, IN 47907, USA
\item[\psinst] Paul Scherrer Institut, PSI, CH-5232 Villigen, Switzerland
\item[\zeuthen] DESY, D-15738 Zeuthen, Germany
\item[\eth] Eidgen\"ossische Technische Hochschule, ETH Z\"urich,
     CH-8093 Z\"urich, Switzerland
\item[\hamburg] University of Hamburg, D-22761 Hamburg, Germany
\item[\taiwan] National Central University, Chung-Li, Taiwan, China
\item[\tsinghua] Department of Physics, National Tsing Hua University,
      Taiwan, China
\item[\S]  Supported by the German Bundesministerium 
        f\"ur Bildung, Wissenschaft, Forschung und Technologie
\item[\ddag] Supported by the Hungarian OTKA fund under contract
numbers T019181, F023259 and T037350.
\item[\P] Also supported by the Hungarian OTKA fund under contract
  number T026178.
\item[$\flat$] Supported also by the Comisi\'on Interministerial de Ciencia y 
        Tecnolog{\'\i}a.
\item[$\sharp$] Also supported by CONICET and Universidad Nacional de La Plata,
        CC 67, 1900 La Plata, Argentina.
\item[$\triangle$] Supported by the National Natural Science
  Foundation of China.
\end{list}
}
\vfill
